\documentclass[preprint,superscriptaddress,footinbib,showkeys,showpacs,tightenlines]{revtex4} 
\usepackage{graphicx}
\begin{document}      
\preprint{PNU-NTG-04/2007}
\preprint{PNU-NURI-04/2007}
\title{A new candidate for non-strangeness pentaquarks: $N^*(1675)$} 
\author{Seung-il Nam}
\email{sinam@yukawa.kyoto-u.ac.jp,sinam@pusan.ac.kr}
\affiliation{Yukawa Institute for Theoretical Physics (YITP), \\Kyoto
University, Kyoto 606-8502, Japan}
\affiliation{Department of
Physics and Nuclear Physics \& Radiation Technology Institute (NuRI),
Pusan National University, Busan 609-735, Republic of Korea} 
\author{Ki-Seok Choi}
\email{kschoi@pusan.ac.kr}
\affiliation{Department of
Physics and Nuclear Physics \& Radiation Technology Institute (NuRI),
Pusan National University, Busan 609-735, Republic of Korea} 
\author{Atsushi Hosaka}
\email{hosaka@rcnp.osaka-u.ac.jp}
\affiliation{Research Center for Nuclear Physics, Osaka University,
Ibaraki, 567-0047 Japan}
\author{Hyun-Chul Kim}
\email{hchkim@pusan.ac.kr}
\affiliation{Department of
Physics and Nuclear Physics \& Radiation Technology Institute (NuRI),
Pusan National University, Busan 609-735, Republic of Korea} 
\date{\today}
\begin{abstract}  
We study a new nucleon resonance from $\eta$ photoproduction,
which was observed at $\sqrt{s}=1675$ MeV with a narrow 
decay width ($\sim 10$ MeV) by the Tohoku LNS group as well as the 
GRAAL collaboration.  Using an effective Lagrangian approach, we
compute differential cross sections for the $\eta$ photoproduction.
In addition to $N^*(1675,1/2^{\pm},3/2^{\pm})$, we employ six other
nucleon resonances, i.e. $N^*(1520,1535,1650,1675,1710,1720)$ and
vector meson exchanges which are the most relevant ones to this
reaction process.  As a result, we can reproduce the GRAAL data
qualitatively well and observe obvious isospin asymmetry between the
transition magnetic moments of $N^*(1675)$: $\mu_{\gamma 
  nn^*}\gg\mu_{\gamma pp^*}$, which indicates that the newly found
nucleon resonance may be identified as a non-strange pentaquark
state. 
\end{abstract}
\pacs{13.75.Cs, 14.20.-c}
\keywords{$\eta$ photoproduction, GRAAL experiment, Pentaquark}
\maketitle
\section{introduction}
Recently, the GRAAL collaboration reported a new nucleon resonance
$N^*(1675)$ from $\eta$
photoproduction~\cite{Kuznetsov:2004gy,Kuznetsov:2006kt}.  Reportedly,  
its decay width $\Gamma_{N^*\to\eta N}$ was estimated to be about $40$
MeV.  However, the Fermi motion being taken into account, its width
may be found to be around 10 MeV~\cite{Fix:2007st}.  This narrow width
is a typical feature for the pentaquark exotic baryons.  Moreover,
the production process of the $N^*(1675)$ largely depends on its
isospin state of the target nucleons: A larger $N^*(1675)$ peak is
shown for the neutron target, while it is suppressed in the proton
one.  Considering the fact that isospin symmetry breaking is
negligible for the strong coupling constants, this large asymmetry
comes mainly from the anomalous electromagnetic couplings,
i.e. $\mu_{\gamma nn^*}\gg\mu_{\gamma pp^*}$.  Interestingly,
$N^*(1675)$ being assumed as a member of the baryon antidecuplet
($\bar{10}$), this large isospin asymmetry was well explained in the
chiral quark-soliton model ($\chi$QSM)~\cite{Kim:2005gz}.  In the
present work, following our previous work~\cite{Choi:2005ki}, we would
like to present a recent study on the $\eta$ photoproduction employing
the effective Lagrangian approach in the Born approximation.  We
include six nucleon resonances ($N^*(1520,1535,1650,1675,1710,1720)$)
as well as usual Born terms as backgrounds.  Theoretically, we assume
for the new $N^*$ resonance both $J^P=1/2^{\pm}$ and 
$3/2^{\pm}$.  We will show in the present report that the large
isospin asymmetry is found in the case of both $J^P=1/2^{\pm}$ and 
$3/2^{\pm}$.  
\section{General Formalism}
First, we define the effective Lagrangians for $\eta$
photoproduction. Since those for the background contributions of spin  
1/2 and 3/2 resonances can be found in the previous 
work~\cite{Choi:2005ki}, we present here the effective Lagrangians
only for the spin 5/2 resonance ($D_{15}$):  
\begin{eqnarray}
{\cal L}_{\eta N N^*}^{5/2}&=&\frac{g_{\eta NN^*}}{M^2_{\eta}}
{N^*}^{\mu\nu}\Theta_{\mu\delta}(X)\Theta_{\nu\lambda}(Y)\Gamma^b_5N 
\partial^{\delta}\partial^{\lambda}\eta+{\rm h.c.}, \nonumber\\
{\cal L}_{\gamma NN^*}^{5/2}
&=&\frac{e\kappa}{M^2_{N^*}}\bar{N}^{*\mu\alpha}\Theta_{\mu\nu}(X)
\gamma_{\lambda}\Gamma^a_5(\partial_{\alpha}F^{\lambda\nu})N+{\rm h.c.},
\end{eqnarray}
where $N$, $N^*$, $\eta$ and $F_{\mu\nu}$ are the fields of the
nucleon, nucleon resonance, pseudoscalar $\eta$ meson and field
strength tensor of the photon, respectively.  $M_h$ stands for the
corresponding mass of hadrons, $h$. The off-shell term can be written
as:   
\begin{equation}
\Theta_{\mu\nu}(X)=g_{\mu\nu}+X\gamma_{\mu}\gamma_{\nu},
\end{equation}
where $X$ is the off-shell parameter which can be determined
phenomenologically.  In the present work, we set it ($X,Y$) to be zero
for convenience.  The Dirac spin matrix $\Gamma^{a,b}_5$ is defined as
follows:  
\begin{eqnarray}
{\pi=+1}&:&\Gamma^a_5={\bf 1}_{4\times4},\,\,\,\,\Gamma^b_5=\gamma_5,
\nonumber\\ 
{\pi=-1}&:&\Gamma^a_5=\gamma_5,\,\,\,\,\Gamma^b_5={\bf 1}_{4\times4} 
\end{eqnarray}
for a given parity $\pi$.  We list the coupling strengths for the
background contributions taken from the Nijmegen potential
model~\cite{Stoks:1999bz} in the table below:  
\begin{table}[h]
\begin{center}
\begin{tabular}{|c|c|c|c|c|c|c|}
\hline
$g_{\eta NN}$&$g_{\rho NN}^v$&$g_{\rho NN}^t$&
$g_{\omega NN}^v$&$g_{\omega NN}^t$&$g_{\rho \eta \gamma}$&$g_{\omega 
  \eta \gamma}$\\ 
\hline
0.47&2.97&12.52&10.36&4.20&0.89&0.192\\
\hline
\end{tabular}
\end{center}
\label{table0}
\end{table}

To determine the anomalous couplings $\mu_{\gamma NN^*}$ and strong
coupling constant $g_{\eta NN^*}$, we use the following equations for
the spin 5/2 resonances: 
\begin{eqnarray}
 \kappa &=& \mp S_a^{5/2}\frac{\sqrt{5}C}{2\sqrt{2M_{N^*}+M^2_N}}
\Big(\frac{M_{N^*}}{k^*}\Big)^2[(A_a^{1/2})^2+(A_a^{3/2})^2]^{1/2},\\
  \nonumber\\
g_{\eta N N^*}&=&\sqrt{\frac{30\pi M_{N^*}M_{\eta}^4
\Gamma_{N^* \to\eta N}}{|\vec{P}_f|^5(\sqrt{M^2_N+k^2}\pm M_N)}}.
\end{eqnarray}
For more details, one can refer to Ref.~\cite{Choi:2007}.  In
Table~\ref{table2} we show the branching ratios and helicity
amplitudes for the resonances to compute the stregths of the
coupling constants.  Note that our choices for those values are  
all within the experimental ones shown in the parentheses.   
\begin{table}[b]
\begin{tabular}{|c||c|c|c|c|c|}
\hline
\hline
$N^*$&$\Gamma_{N^*\to\eta N}/\Gamma_{N^*}$ [$\%$]&$A^n_{1/2}$
[GeV$^{-1/2}$]& $A^p_{1/2}$ [GeV$^{-1/2}$]&$A^n_{3/2}$
[GeV$^{-1/2}$]&$A^p_{3/2}$ [GeV$^{-1/2}$]\\ 
\hline
\hline
$1520$&$2.3\times 10^{-3}$&$-0.059$&$-0.024$&$-0.139$&$0.166$\\
&($2.3\pm 0.4 \times
10^{-3}$)&($-0.059\pm0.009$)&($-0.024\pm0.009$)&($-0.139\pm0.011$)
&($0.166\pm0.005$)\\  
\hline 
$1535$&55
&$-0.060$&0.085 &-&-\\
&($45\sim60$)&($-0.046\pm0.027$)&($0.090\pm0.030$)&&\\
\hline
$1650$&7&$-0.020$&0.060&-&-\\
&($3\sim10$)&($-0.015\pm0.021$)&($0.053\pm0.016$)&&\\
\hline
$1675$&$1$&$-0.031$&$0.011$&$-0.045$&$0.006$\\
&($0.0\pm1$)&($-0.043\pm0.012$)&($0.019\pm0.008$)&($-0.058\pm0.013$)
&($0.015\pm0.009$)\\
\hline
$1710$&6 &$-0.002$&0.009 &-&-\\
&($6\pm1$)&($-0.002\pm0.014$)&($0.009\pm0.022$)&&\\
\hline
$1720$&4 &0.001
&0.001 &$-0.029$
&$-0.010$ \\
&($4\pm1$)&($0.001\pm0.015$)&($0.018\pm0.030$)&($-0.029\pm0.061$)
&($-0.019\pm0.020$)\\
\hline
\hline
\end{tabular}
\caption{The parameters of the nucleon resonances : Branching ratios
and helicity amplitudes for the neutron and proton.}  
\label{table2}
\end{table} 
Considering the spatial distributions for the hadrons involved, we
employ four dimensional relativistic form factors in a
gauge-invariant manner: 
\begin{equation}
F^h_x=\frac{\Lambda^4}{\Lambda^4+(x-M^2_h)^2},
\end{equation}
where $x$ is the square of transfered momentum (Mandelstam
variable).  The cut off masses are determined as follows: 
\begin{equation}
\Lambda_{N^*}=1.4\,{\rm GeV},\,\,\Lambda_{N}=0.8\,{\rm GeV}
,\,\,\Lambda_{\rho}=1.1\,{\rm GeV},\,\,\Lambda_{\omega}=1.1\,{\rm GeV}.
\end{equation}

\section{Numerical results}
\begin{figure}[h]
\begin{tabular}{cccc}
\includegraphics[width=3.5cm]{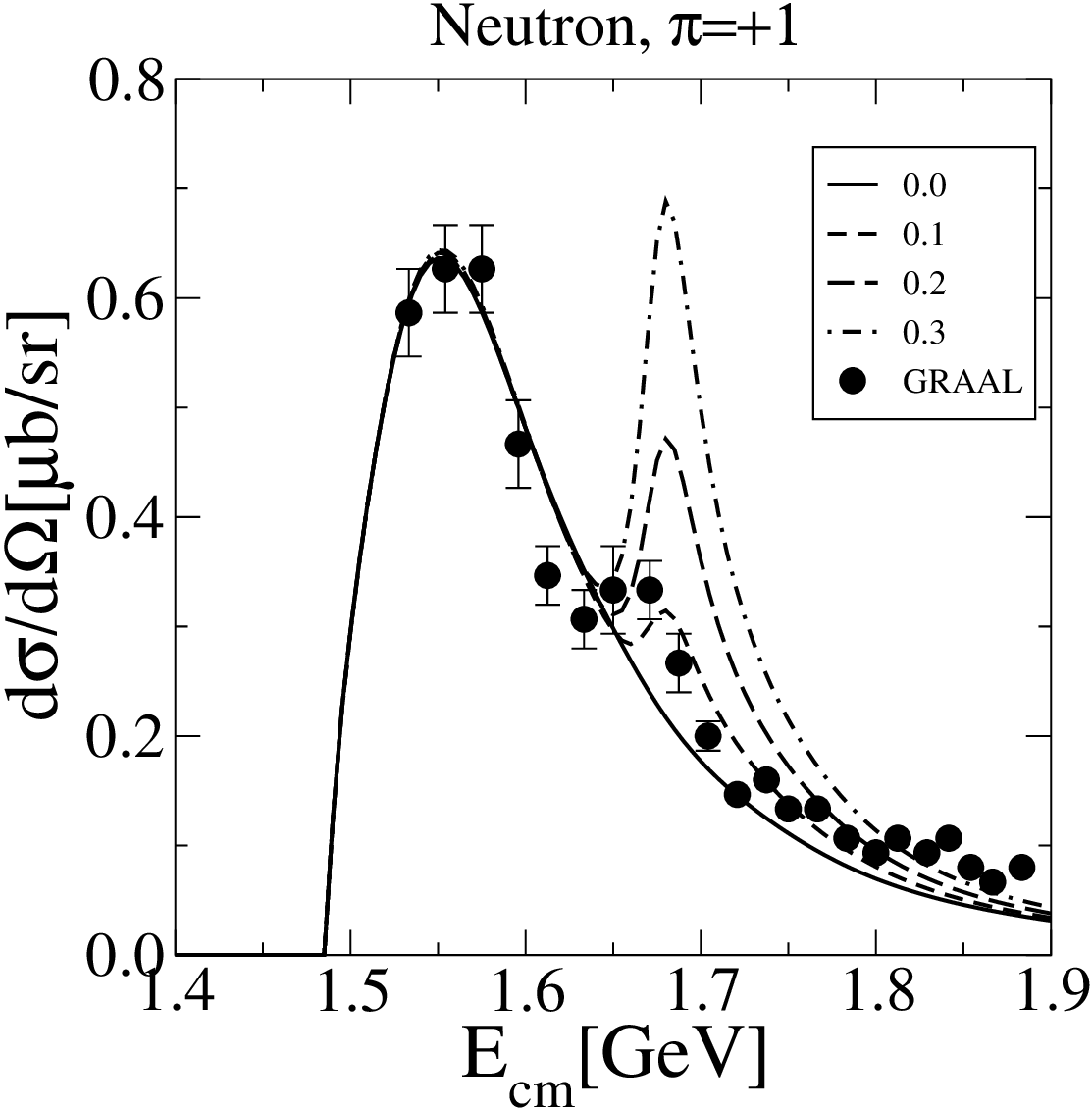}
\includegraphics[width=3.5cm]{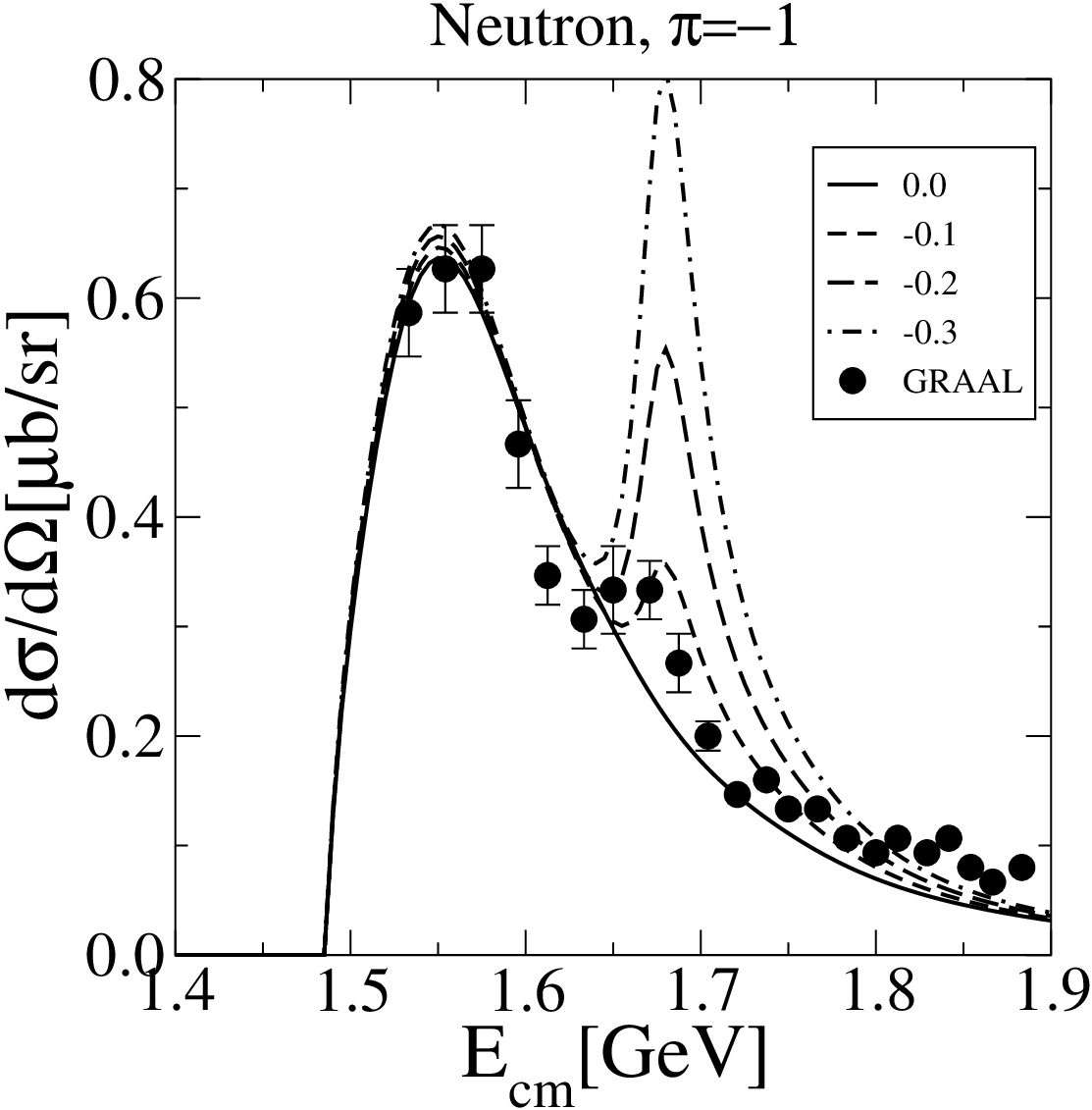}
 \includegraphics[width=3.5cm]{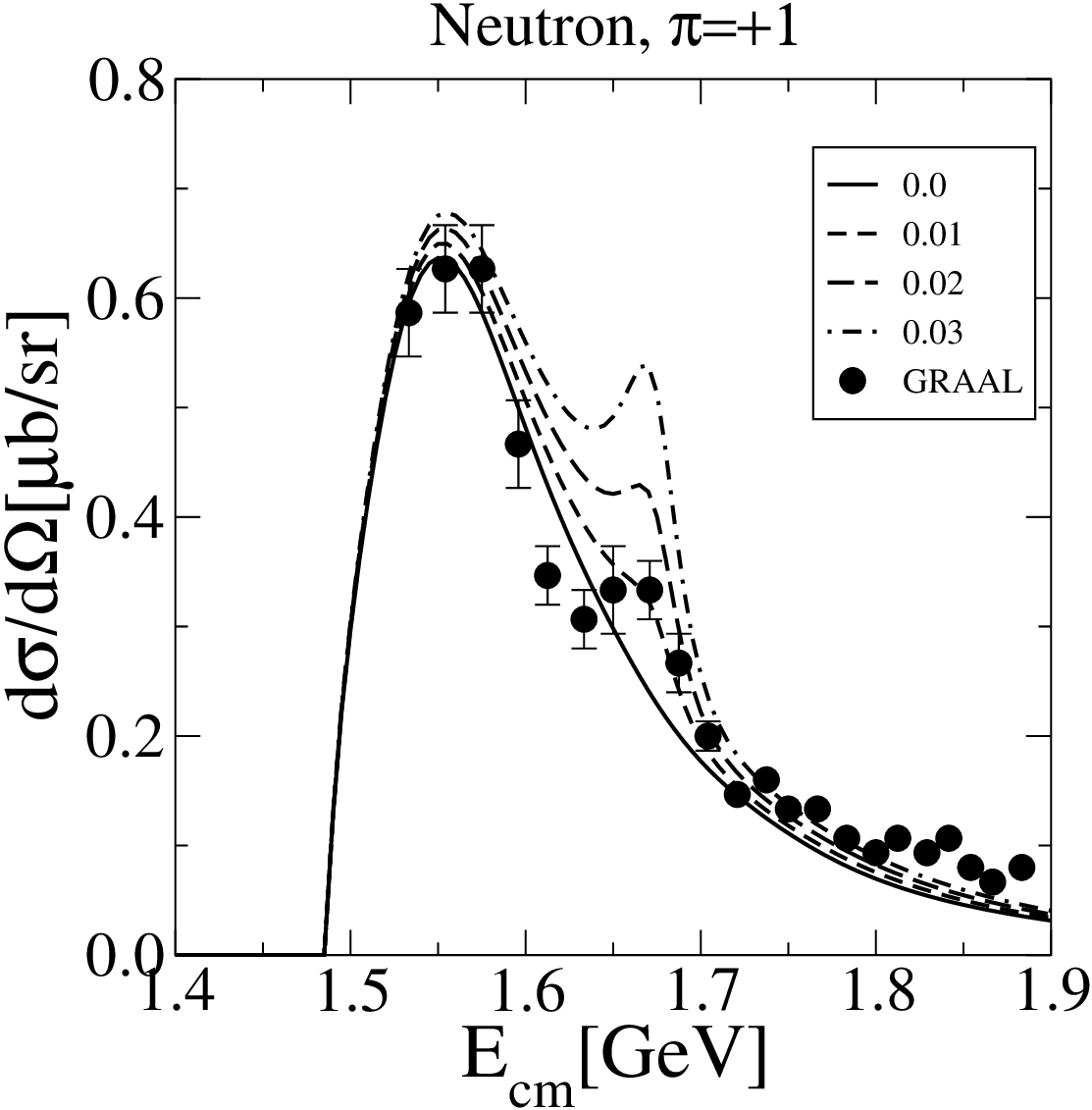}
\includegraphics[width=3.5cm]{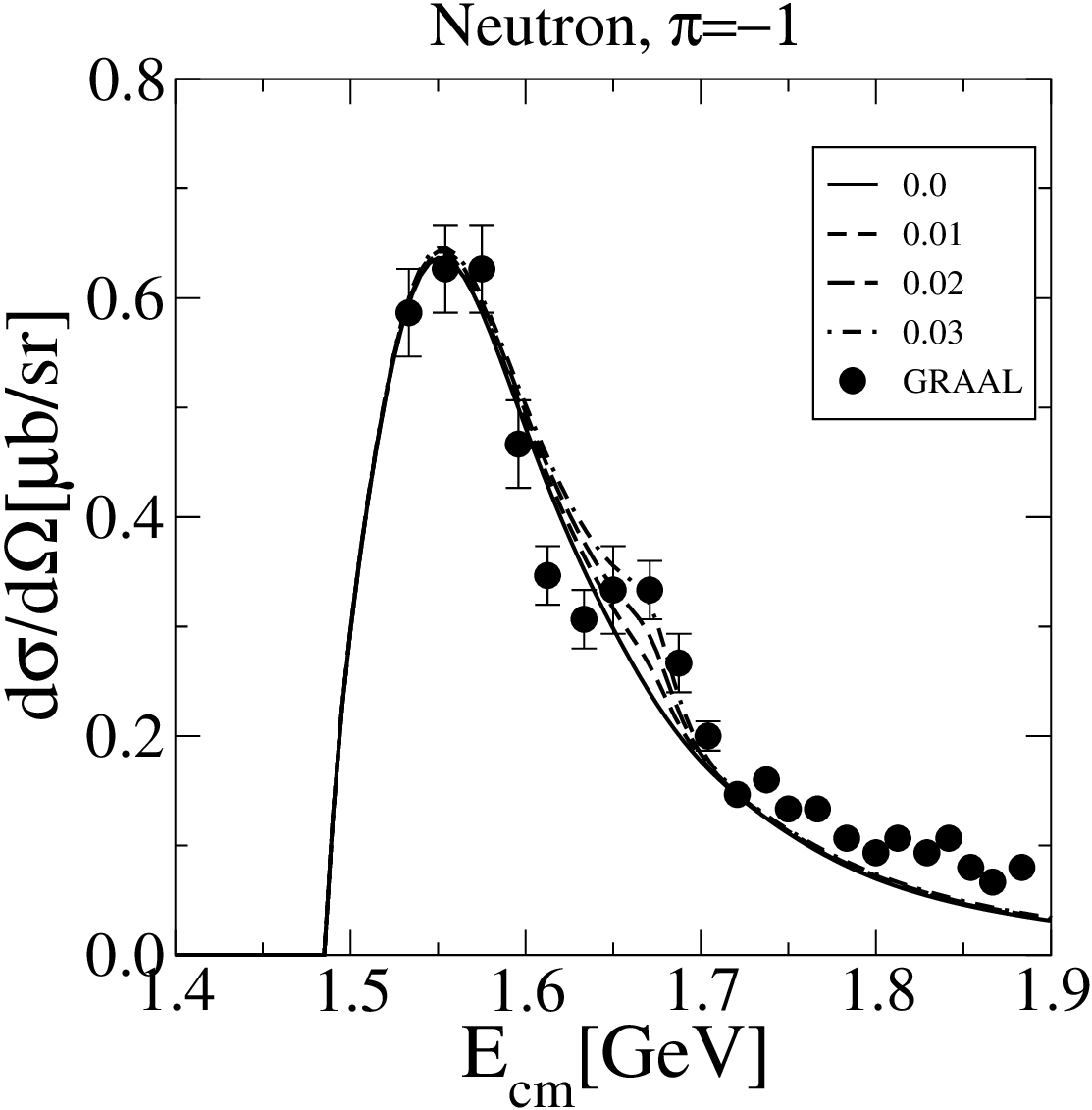}
\end{tabular}
\caption{Differential cross sections for the neutron target. From the left
to the right, we show the results of $N^*(1675)$ for $J^P=1/2^+$, $1/2^-$,
$3/2^+$, and $3/2^-$, respectively.} 
\label{fig1}
\end{figure}

\begin{figure}[h]
\begin{tabular}{cccc}
\includegraphics[width=3.5cm]{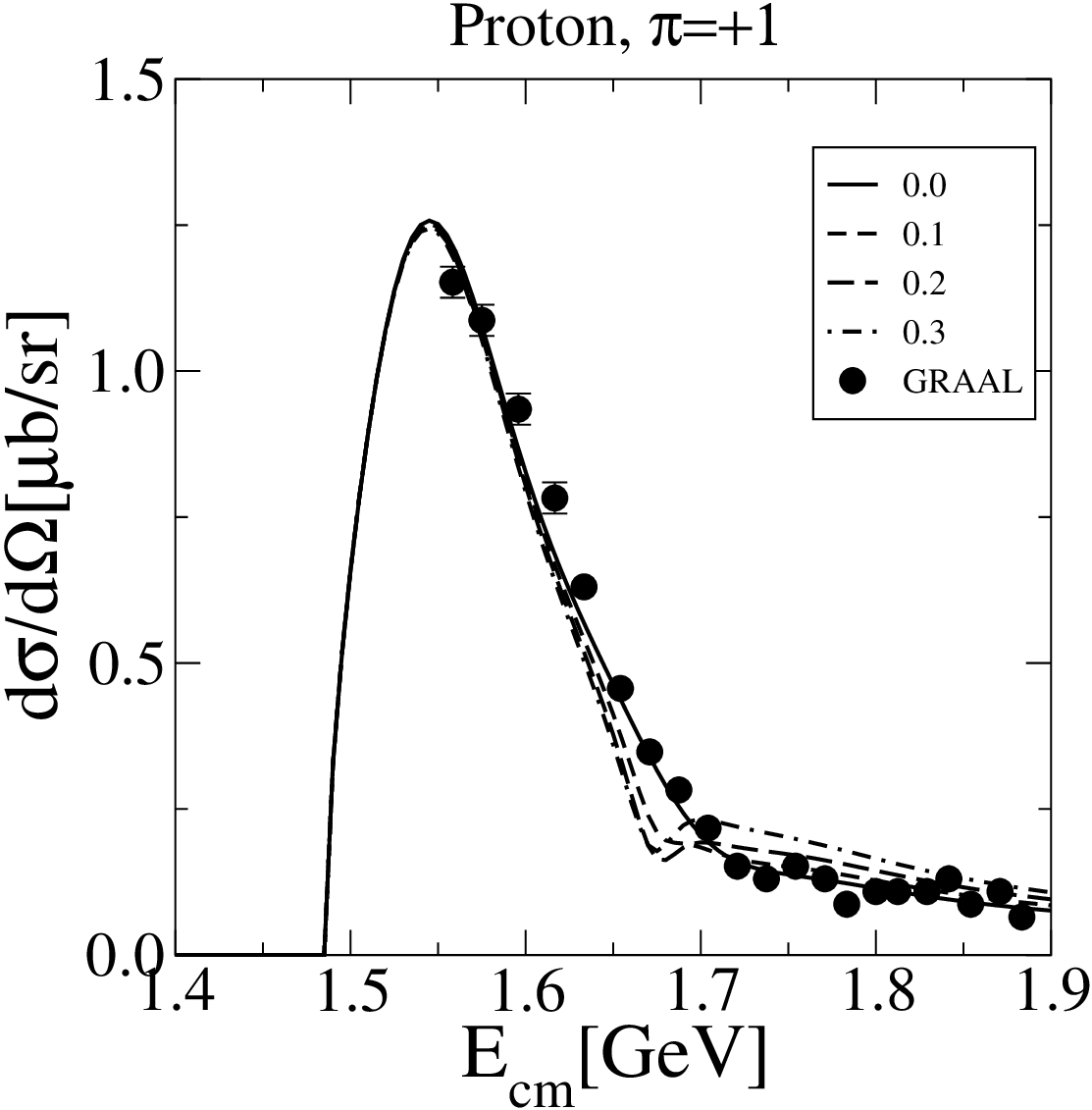}
\includegraphics[width=3.5cm]{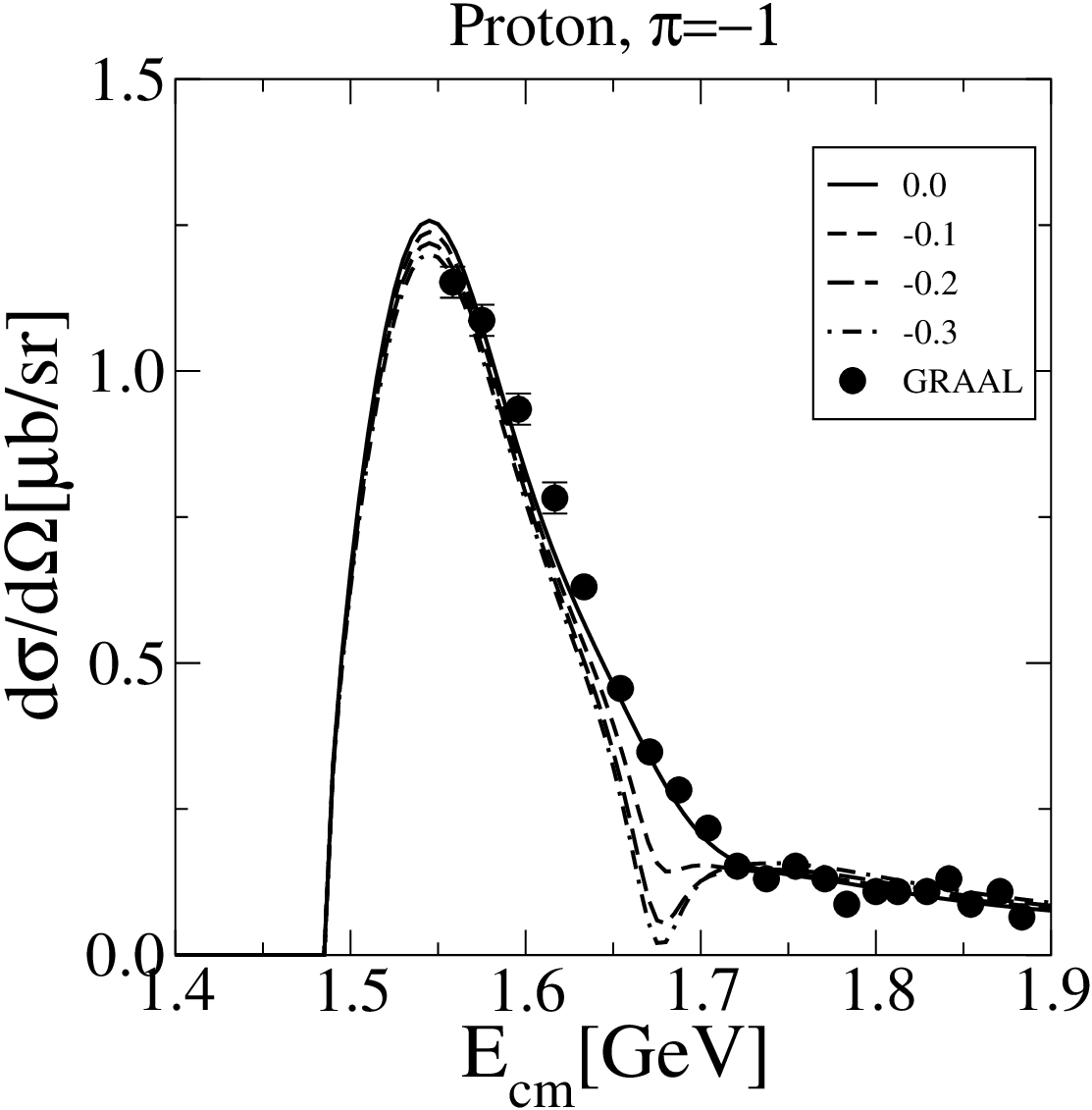}
 \includegraphics[width=3.5cm]{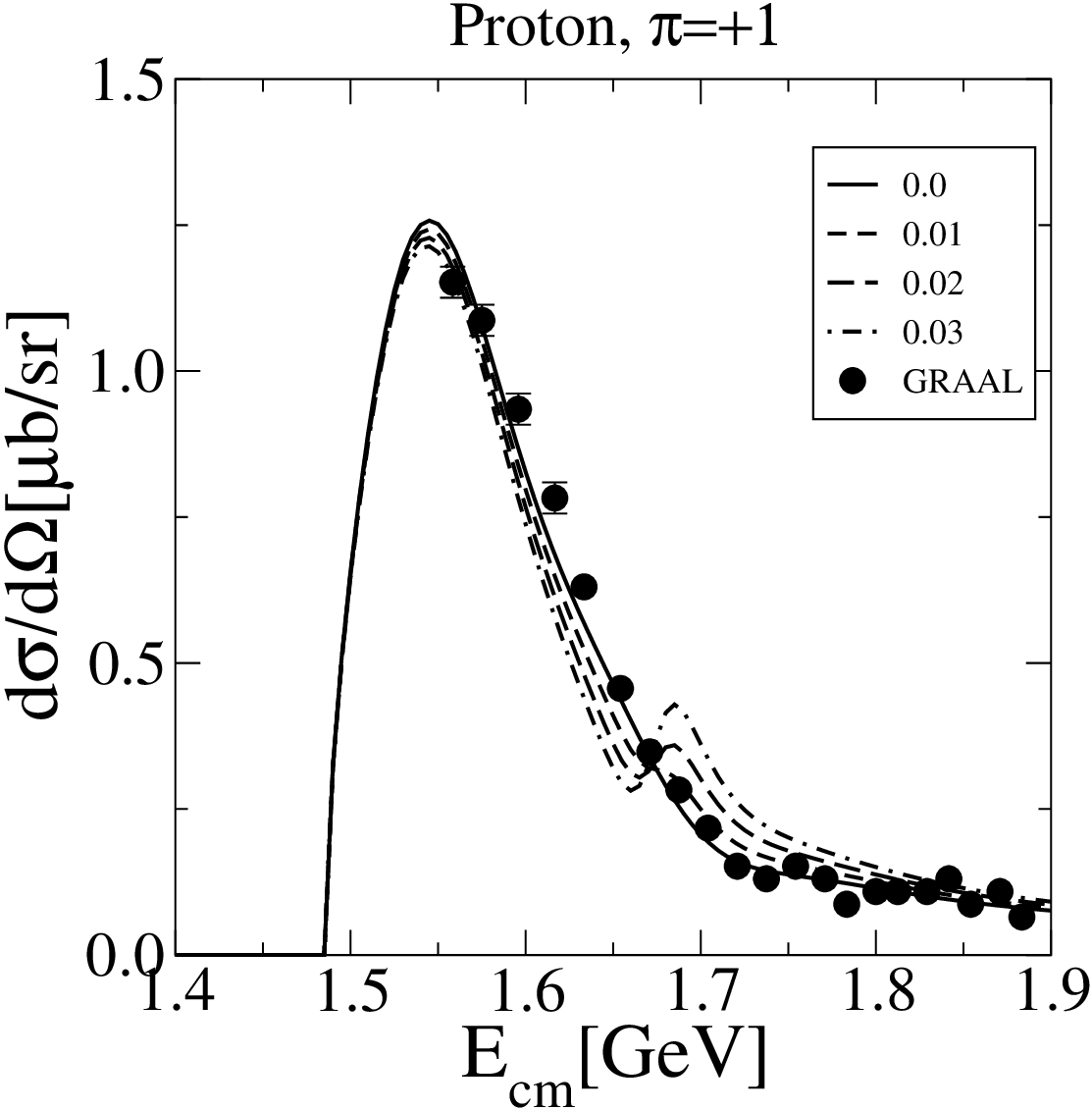}
\includegraphics[width=3.5cm]{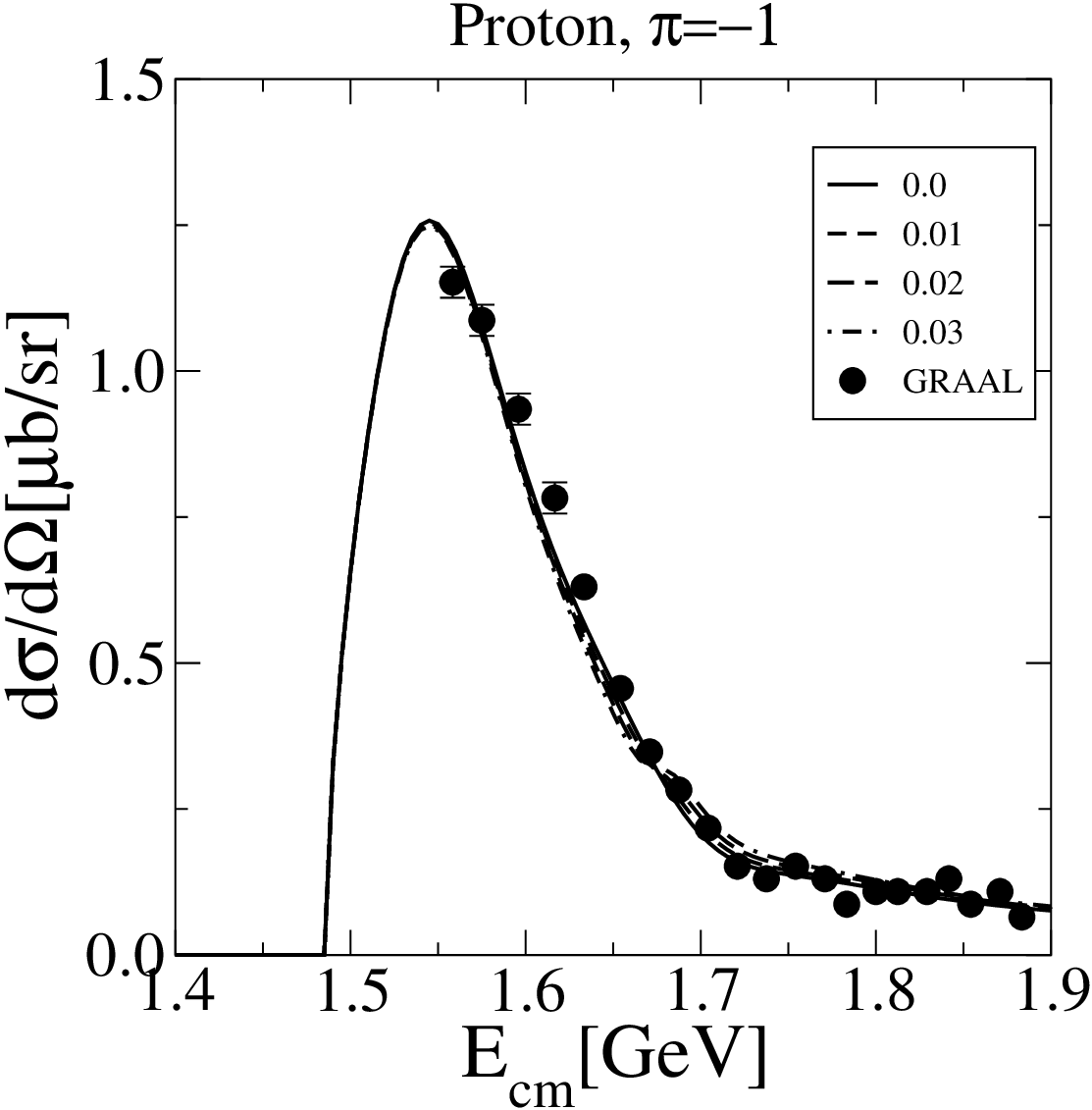}
\end{tabular}
\caption{Differential cross sections for the proton target. From the
  left to the right, we show the results of $N^*(1675)$ for
  $J^P=1/2^+$, $1/2^-$, $3/2^+$ and $3/2^-$, respectively.} 
\label{fig2}
\end{figure}

We focus here on the differential cross sections which were measured
by the GRAAL experiment. In Figure~\ref{fig1}, we show the results for
the neutron targert, assuming various quantum numbers for the
$N^*(1675)$.  From the left to the right, we show each case for
$J^P=1/2^+$, $1/2^-$, $3/2^+$, and $3/2^-$, respectively.  The 
comparable results to the GRAAL data are obtained for the
$J^P=1/2^{\pm}$ cases, with $|\mu_{\gamma nn^*}|\simeq0.1\sim0.2$
used.   For the spin 3/2 cases, it is rather difficult to see clear
peak structures.  Moreover, the strength of $|\mu_{\gamma nn^*}|$ to
produce the data should be about ten times smaller than that for 
the spin 1/2 cases. The reason for this can be understood by the fact
that higher partial-wave contributions play an important role in the
case of the spin 3/2 $N^*$.     
In Figure~\ref{fig2}, we present similarly the results for the proton
target.  As for all the cases, the strength of $|\mu_{\gamma pp^*}|$
should be very small ($\sim$0) in order to reproduce the data.  
Thus, we find that there exists clearly the isospin asymmetry for the 
$N^*(1675)$ resonance from the $\eta$ photoproduction.  This
observation also gives us a clue that the new nucleon resonance may be
identified as a memeber of the baryon antidecuplet as suggestd in the
$\chi$QSM, although the negative parity was not
explored~\cite{Kim:2005gz}.   

\section{Conclusion}
We have investigated the production mechanism of the newly 
found nucleon resonance $N~*(1675)$ from the $\eta$ photoproduction by
the GRAAL collaboration.  Employing the effetive Laggrangian approach
based on the Born approximation, we computed its different cross
sections for the neutron as well as the proton targets. We observed
a siginificant isospin asymmetry between the anomalous couplings for 
$\gamma NN^*(1675)$ vertex: $|\mu_{\gamma nn^*}|=0.1\sim0.2$ and
$\mu_{\gamma pp^*}\sim0$ for various quantum states of the
$N^*(1675)$. More extensive study will appear
elsewhere~\cite{Choi:2007}.        
\section*{Acknowledgments}
The present work is supported by the Korea Research Foundation Grant
funded by the Korean Government(MOEHRD) (KRF-2006-312-C00507). The works of
S.i.N. and K.S.C. are partially supported by the 
Brain Korea 21 (BK21) project in Center of Excellency for Developing Physics
Researchers of Pusan National University, Korea. The authors thank the
Yukawa Institute for Theoretical Physics at Kyoto University, where
this work was completed during the YKIS2006 on "New Frontiers on QCD". 

\section{References}

\end{document}